# The contribution of the modern amateur astronomer to the science of astronomy

## Filipp Romanov

Amateur astronomer, Russia; member of the AAVSO (American Association of Variable Star Observers). filipp.romanov.27.04.1997@gmail.com ORCID iD: 0000-0002-5268-7735

## Abstract

An amateur astronomer in the modern world has the opportunity not only to make visual observations for own interest, but can make scientific astronomical observations and new discoveries in astronomy.

In my example, as amateur astronomer and only through self-education, I inform about my discoveries: of the possible dwarf nova on the old digitized photographic plates and of new variable stars from sky surveys data by means of data mining; how I discovered (in the images of the sky surveys): astronomical transients, supernovae, planetary nebula candidates and new binary systems in the data of Gaia DR2; I describe my discoveries of three novae in the Andromeda Galaxy.

I report about some of my scientific observations using remote telescopes: of superhumps of cataclysmic variable stars; of echo outburst of AM CVn star; of maximum brightness of blazars; of optical afterglows of gamma-ray bursts (including GRB 221009A); of microlensing events; of rotation of near-Earth asteroid 2022 AB. I also describe my photometric follow-up observations of novae (including V1405 Cas and V1674 Her) and my astrometric observations of Solar System objects (including the confirmation of objects posted at the Confirmation Pages of the Minor Planet Center) including observations of comet 2I/Borisov, asteroids 2020 AV2 and (65803) Didymos. I also describe some of my observations of occultations: of the star by asteroid (159) Aemilia, of the star by Saturn's moon Titan and of Uranus by the Moon during total lunar eclipse on November 8, 2022; and visual observations of variable stars, meteors and sunspots (including during the transit of Venus in 2012).

Some of my data already used in scientific papers, others were sent to the databases. I share my experience of discovery and research of astronomical objects and in my example, I show that an amateur astronomer can make a real contribution to the science.

# 1. Introduction

I have not studied at the university yet (but I want and plan to study astronomy at university in the near future), but I contribute to the science, as an amateur astronomer (I started studying astronomy on August 17, 2009, only on the basis of self-education; before that, for the first time I observed an astronomical phenomenon with the unaided eye on August 28, 2007: the total lunar eclipse), with my astronomical observations and discoveries. I briefly report the scientific data that I have made as an amateur astronomer and the results, showing that they are a real contribution to the science.

## 2. Astronomical discoveries
## 2.1. Analysis of archival sky images

This way I discovered several variable stars with the possible dwarf nova (Romanov V1) on two digitized photographic plates of the Palomar Observatory Sky Survey taken in April 1958. My paper about this: 2018OEJV..190....1R (*Romanov 2018*). **Now I am also guessing that it could be a heavily reddened Galactic nova.**

I also found two possible missed supernovae in digitized photographic plates: AT 1991bm in the galaxy UGC 11180 (details: *Romanov 2019* 2019TNSTR2353....1R) and AT 1992bw in the galaxy UGC 43 (*Romanov 2019* 2019TNSTR2388....1R).

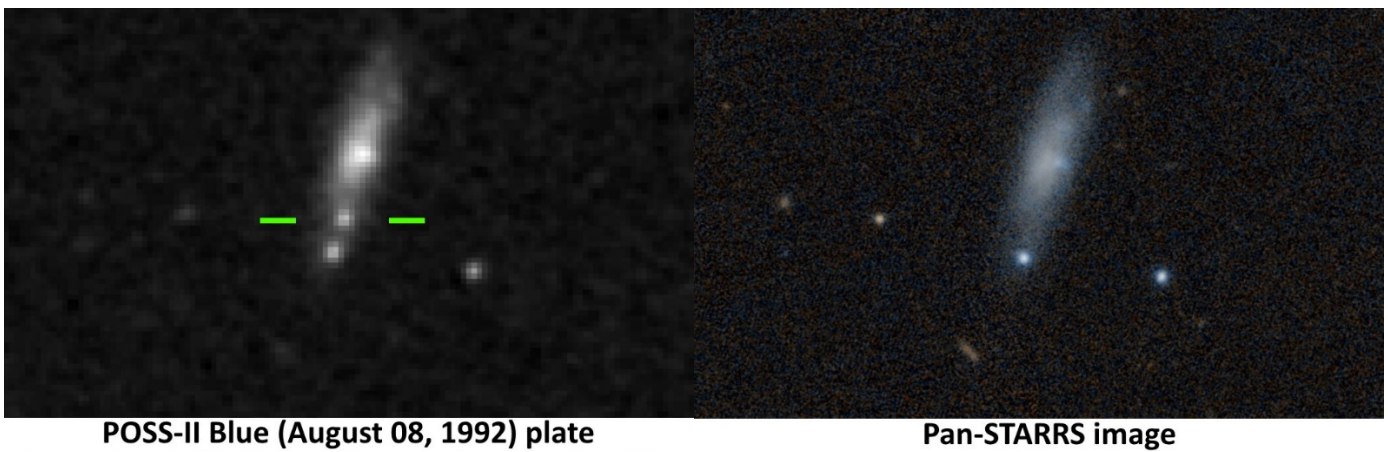

**POSS-II Blue (August 08, 1992) plate**          **Pan-STARRS image**

*Figure 1: AT 1992bw in UGC 43 in the digitized plate compared to Pan-STARRS1 image*
*(Chambers et al. 2016 2016arXiv161205560C)*

I discovered 10 planetary nebula candidates (and 5 independent co-discoveries) comparing the AllWISE images and photos from the sky surveys, mostly SuperCOSMOS H-alpha Survey (SHS). Example: Ro 2 (J2000.0 position: 15:35:25.53, -57:40:34.66): in the figure 2. Information about the first part of my discoveries of PN candidates was published in the VizieR Online Data Catalog: J/other/LAstr/125.54 (*Le Du 2019*).

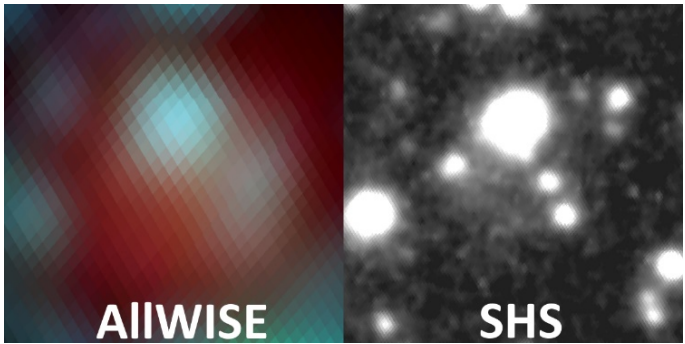

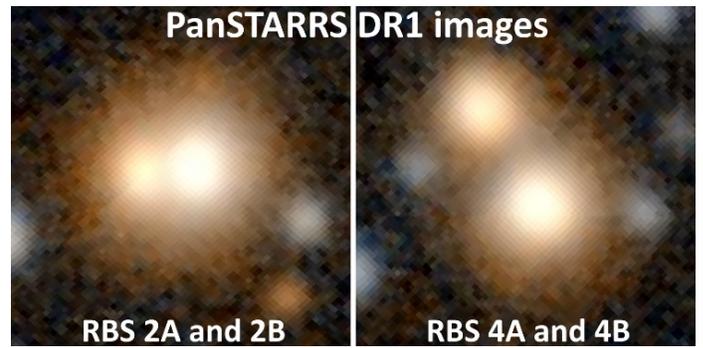

*Figure 2: PN candidate Ro 2 in the AllWISE ([Wright et al. 2010](#) [2010AJ....140.1868W](#)) and SHS ([Parker et al. 2005](#) [2005MNRAS.362..689P](#)) images, FoV ≈ 50 arcseconds*

*Figure 3: Pan-STARRS1 images of pairs of stars RBS 2A with 2B and RBS 4A with 4B, FoV ≈ 15 arcseconds*

In my paper [2019JDSO...15..434R](#) (*Romanov 2019*) I described my discoveries of 4 new pairs of stars (two pairs shown in the figure 3) with almost identical proper motions and parallaxes (probable binary systems) from the Gaia DR2 data. I found some of them when I saw the changing their positions in comparison of images for different years. I also created the VizieR Online Data Catalog [J/other/JDSO/15.434](#).

## 2.2. Analysis of recent sky photographs

I discovered faint (≈ 21.5 mag., confirmed photometrically) supernova candidate AT 2020quo [2020TNSTR2387....1R](#) (*Romanov 2020*) in the Pan-STARRS images during the International Asteroid Search Campaign (I also found the asteroid [2020 OZ7](#) during this). I also discovered supernovae (confirmed by spectra) in the images of the Catalina Real-Time Transient Survey (in the CRTS Great Supernova Hunt project): 91bg-like Type Ia SN 2022bsi in the galaxy NGC 5902 (details: *Romanov and CRTS 2022* [2022TNSTR.339....1R](#)) and SN 2022jhn (low-luminosity Type IIP supernova) in the galaxy PGC 65131 = MCG -01-52-016 (*Romanov and CRTS 2022* [2022TNSTR1171....1R](#)). I am co-author of the classification reports for these supernovae: [2022TNSCR.553....1F](#) (*Fremling, Romanov and ZTF 2022*) and [2022TNSCR1273....1R](#) (*Romanov and Pastorello 2022*) respectively.

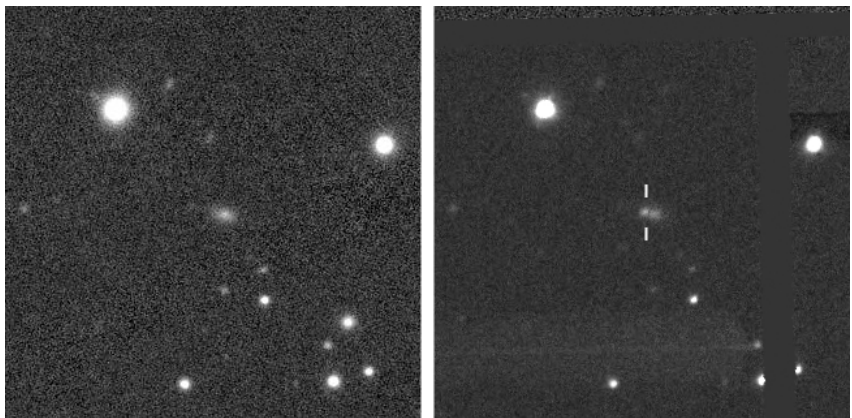

*Figure 4: Pan-STARRS images of the galaxy AllWISE J215216.24-190130.8: before and after the appearance of AT 2020quo*

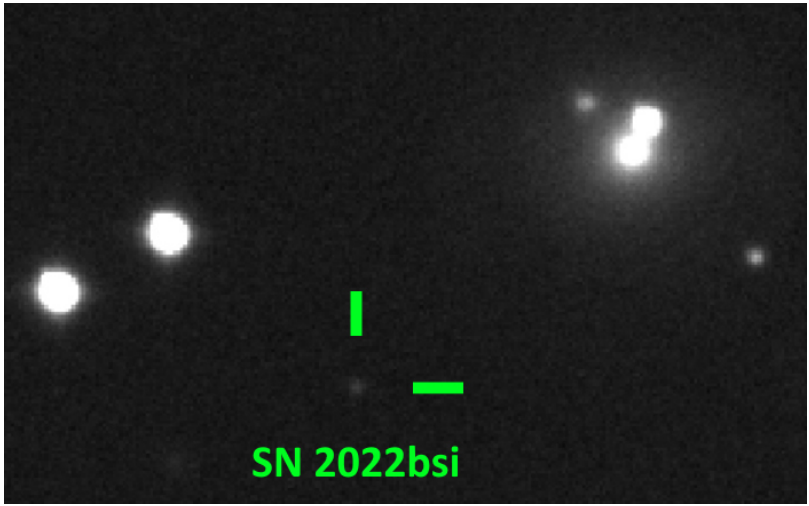

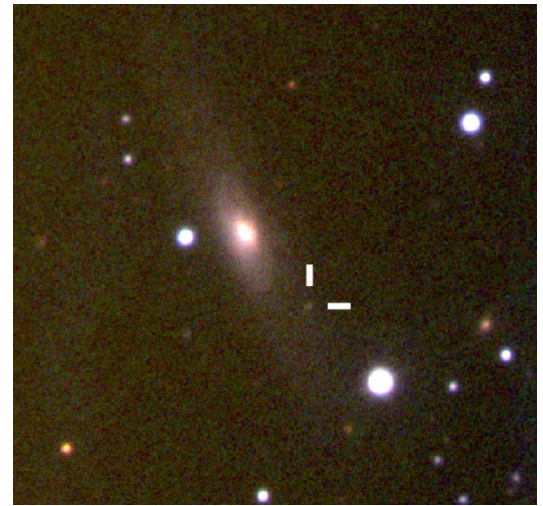

*Figure 5: SN 2022bsi imaged on 2022-02-20 11:46 UT using 0.61-m f/6.5 reflector T24 of iTelescope.Net in Sierra Remote Observatory, Auberry, USA (MPC code U69): stacked 9x300 sec., Luminance filter*

*Figure 6: SN 2022jhn imaged at the Liverpool Telescope on 2022-05-09: stacked 3 x 3x60 sec. with Sloan i', r' and g' filters for red, green and blue channels*

On 2022-07-14 I discovered the apparent nova in M31 named AT 2022ouu = PNV J00423928+4115169: I studied the photographs obtained (after my requests) remotely using 0.61-m f/6.5 public robotic telescope (*Lane 2018* 2018RTSRE...1..119L) of the Burke-Gaffney Observatory (BGO) located in Halifax, Nova Scotia, Canada (MPC code 851). Details: 2022TNSTR1975....1R (*Romanov 2022*). The spectrum (presented in the figure 7) was obtained during the IAUGA 2022, using the low resolution (approximately R = 350 = 18 Å) SPRAT spectrograph on the 2-m f/10 robotic Liverpool Telescope (La Palma, Spain; MPC code J13; *Steele et al. 2004* 2004SPIE.5489..679S): I classified it as Fe II nova, detailed information is available in 2022TNSCR2408....1R (*Romanov 2022*) and in the ATel #15569 2022ATel15569....1R (*Romanov 2022*).

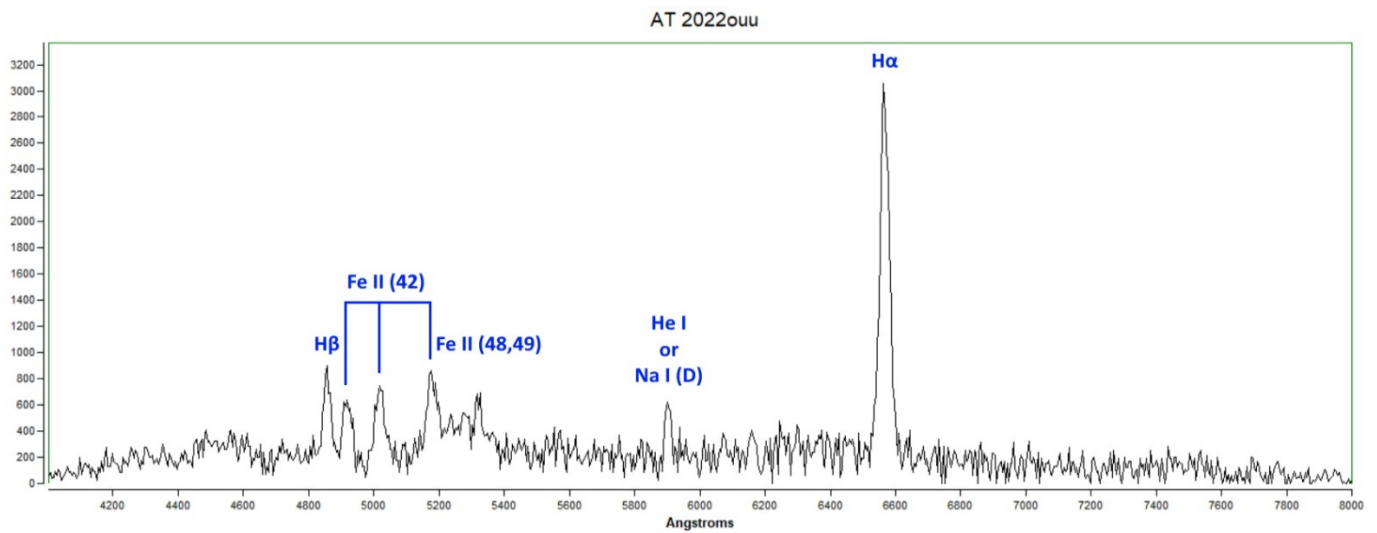

*Figure 7: spectrum of AT 2022ouu obtained during 1800 seconds on 2022-08-04, midtime 03:58:02 UT. FWHM of the Hα line ≈ 34 Å, corresponding to the velocity of ≈ 1550 km/s*

Also during the IAUGA 2022 I discovered two more novae in M31: AT 2022qug = PNV J00424712+4117103 on 2022-08-07 (*Romanov 2022* 2022TNSTR2248....1R) at the images from telescope of Burke-Gaffney Observatory and AT 2022qzf = PNV J00420940+4115311 on 2022-08-09 (*Romanov 2022* 2022TNSTR2276....1R) at the images from telescope T24 of iTelescope.Net. They respectively were confirmed by the spectra obtained on 2022-08-22 at 1.82-m Copernico Telescope of Mt. Ekar near Asiago (*Ochner et al. 2022* 2022TNSCR2436....1O) and on 2022-08-14 at the Palomar 1.5-m telescope (*Perley et al. 2022* 2022TNSCR2345....1P; ZTF designation of the nova AT 2022qzf is ZTF22aazmooy).
**I conclude from the published spectra that these novae are also Fe II novae in M31.**

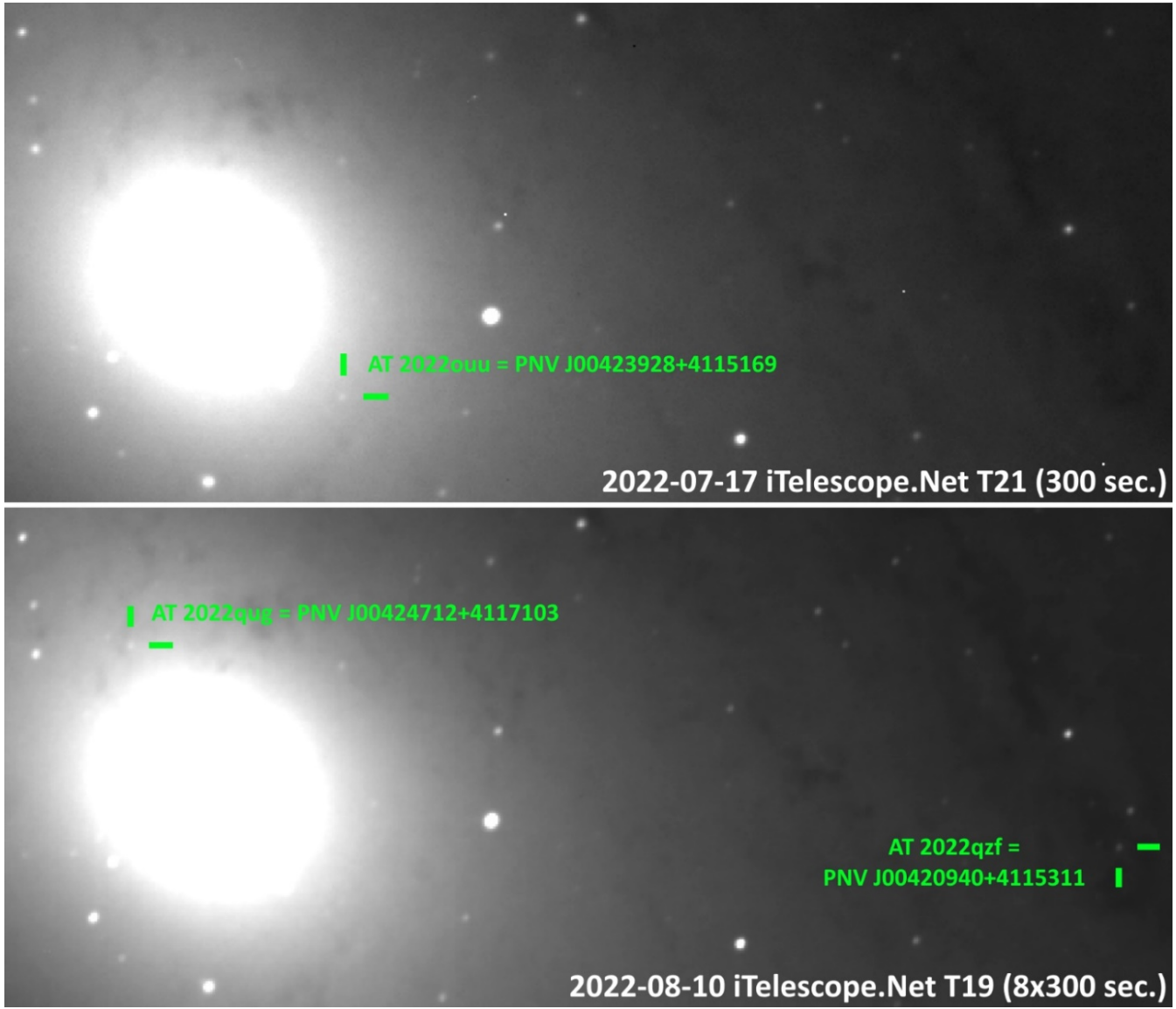

*Figure 8: images of three novae in M31 discovered by me in the summer of 2022. Photographed according to my requests using iTelescope.Net remote telescopes: T21 (0.43-m f/6.8 reflector + CCD + f/4.5 focal reducer) and T19 (0.43-m f/6.8 reflector + CCD) located in New Mexico Skies at Mayhill, New Mexico, USA (MPC code H06)*

## 2.3. Discoveries of variable stars using data mining

I discovered 80 variable stars (the first was a semiregular NSVS 3246176 = IRAS 20414+5244 in January 2016; its J2000.0 position is 20:42:54.90, +52:55:35.00 from the Gaia DR2 catalog; in 2017 I found: periods of 38.96 and 27.12 days and magnitude range 13.15 - 13.50 V), most of them I found as a result of the analysis of photometric data from several sky surveys, information about them has been added to the AAVSO VSX 2006SASS...25...47W (*Watson et al. 2006*). In my paper 2021JAVSO..49..130R (*Romanov 2021*) published in the JAAVSO, I informed about the discovery (using ADQL) and research of eclipsing variable star Romanov V20 (I presented this discovery at the 109th Annual Meeting of the AAVSO which was held online on November 13-15, 2020).

Another example: I found variable star Romanov V48, it may be an intermediate object between intermediate polars and polars, this is reported in the paper 2022arXiv220402598K (*Kato and Romanov 2022*).

## 3. Scientific observations using remote telescopes (located at different observatories, in different hemispheres of the Earth)

I remotely request time series of images of some cataclysmic variable stars during their outbursts, then I do the photometric measurements (including for study the amplitudes and periods of superhumps), I send data to the AAVSO (my Observer Code is RFDA) and to VSNET. Examples:

### 3.1. Cataclysmic variable stars

Figure 9 shows the V-band light curve (time is indicated in Julian days - JD) of eclipsing UGSU-type dwarf nova HT Cas based on images obtained (at my request) on 2021 June 11, remotely at 0.355-m f/6.2 telescope of Abbey Ridge Observatory, Canada (the owner of the observatory David James Lane gave me permission for such imaging).

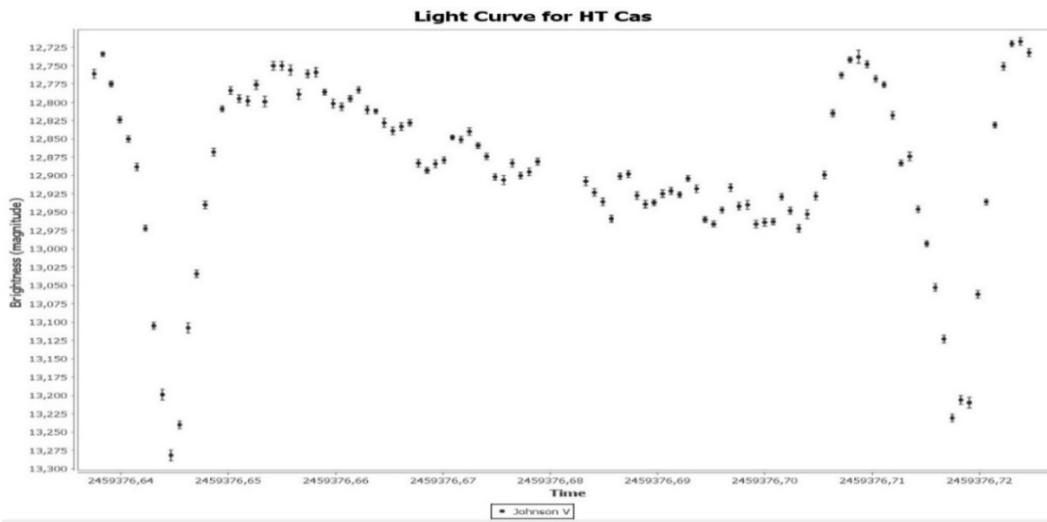

*Figure 9*

Figure 10 shows the JD light curve (with V zero point) showing superhumps of variable star PNV J03022732+1917552 = AT 2021afpi (WZ Sge-type dwarf nova having the largest outburst amplitude known to date, according to *Isogai et al. 2021* 2021ATel15074....1I) based on my data measured from unfiltered images taken remotely at the telescope of Abbey Ridge Observatory (ARO; MPC code I22) during 3 hours on the night of January 4/5, 2022.

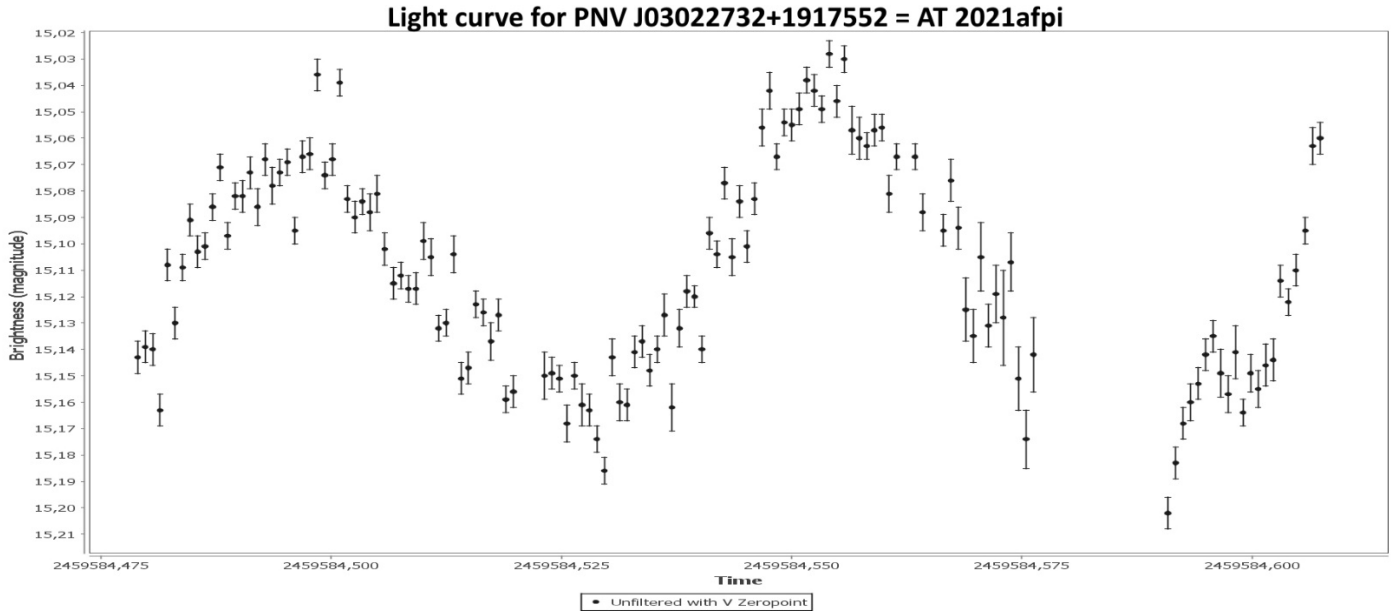

*Figure 10*

On March 4, 2021, I detected the rebrightening (during echo outburst) of cataclysmic variable star ASASSN-21au. In the figure 11: 1st photo was taken at the ARO (mag.= 16.1 V), 2nd photo was taken almost 7 hours later at the BGO (mag.= 14.19 V). Based on my observations, I became a co-author of the paper 2022ApJ...926...10R (*Rivera Sandoval et al. 2022*).

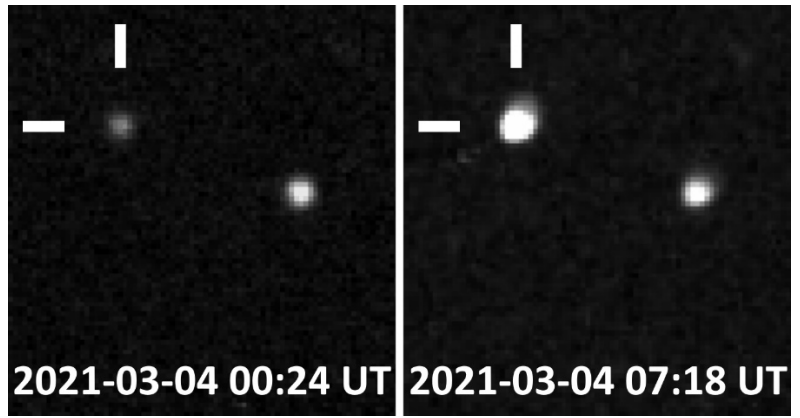

*Figure 11*

On 2022-04-03 I detected re-appearance of optical flickering of RS Oph according to my photometric data from photographs obtained on the BGO telescope with Johnson B filter. It started between March 30 and April 3, 2022, because during observations on 2022-03-30, the mean magnitude was 13.14 B +/- 0.05 and magnitude errors were similar to those of comparison stars. I described it in the ATel #15339 (*Romanov 2022*) 2022ATel15339....1R.

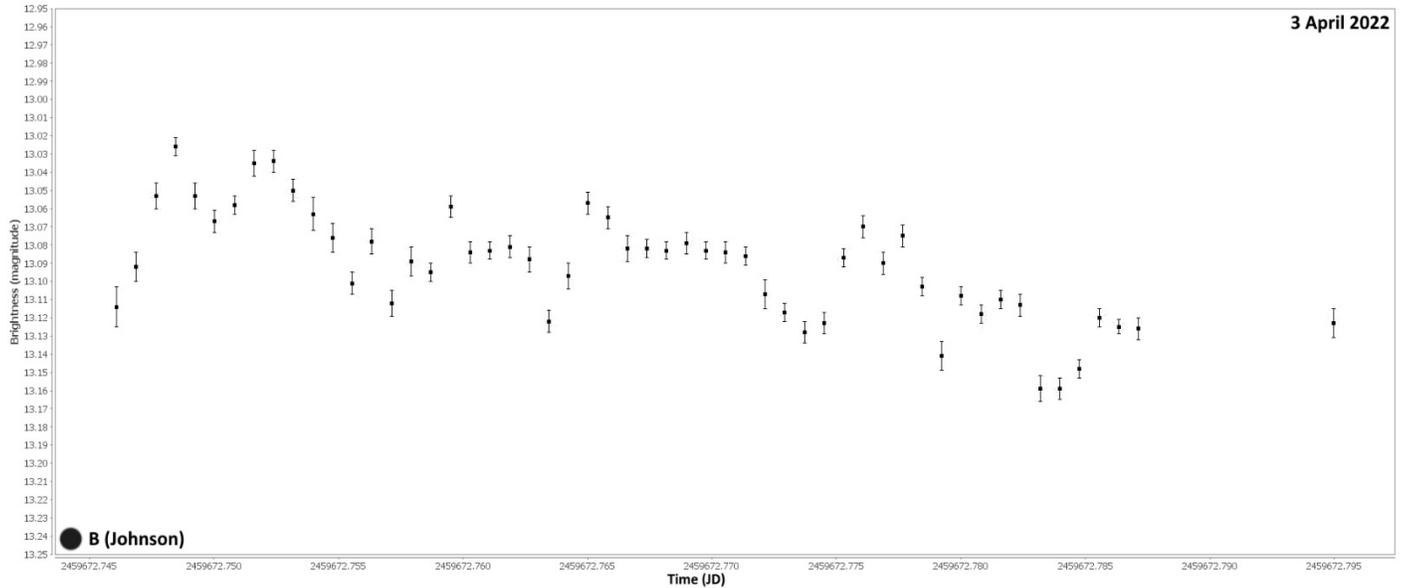

*Figure 12: JD light curve of RS Oph created from data from photographs taken on April 3, 2022.*
*I detected apparent magnitude variations with a peak-to-peak amplitude of 0.13 mag in B band*

I also make follow-up observations (CCD and DSLR) of bright novae that appear on the CBAT "Transient Objects Confirmation Page", some of my photometric data are published in the Central Bureau Electronic Telegrams. Examples: novae V1405 Cas (CBET 4945: *Green 2021*) and V1674 Her (CBET 4976: *Green 2021* and CBET 4977: *Green 2021*). Results of my observations of novae PGIR22gjh, U Sco and FQ Cir (figure 14: photo was taken using iTelescope.Net, from where I was provided with some complimentary points for imaging) were published in the ATel #15511 (*Romanov 2022*) 2022ATel15511....1R.

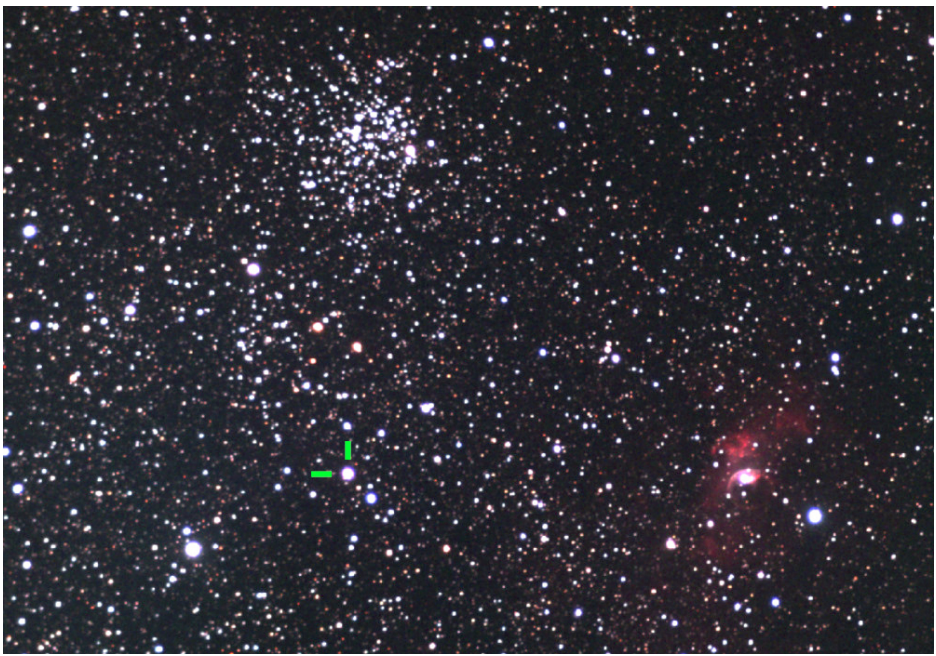

*Figure 13. Nova V1405 Cas (near Messier 52 open cluster of stars and NGC 7635 nebula) imaged (according to my requests) on March 22 and 23, 2021, at 0.1-m f/5 refractor + CMOS of the MRO: Mini-Robotic Observatory (located at Abbey Ridge Observatory, Canada): stacked 4x900 sec. (Red, Green, Blue and Luminance filters)*

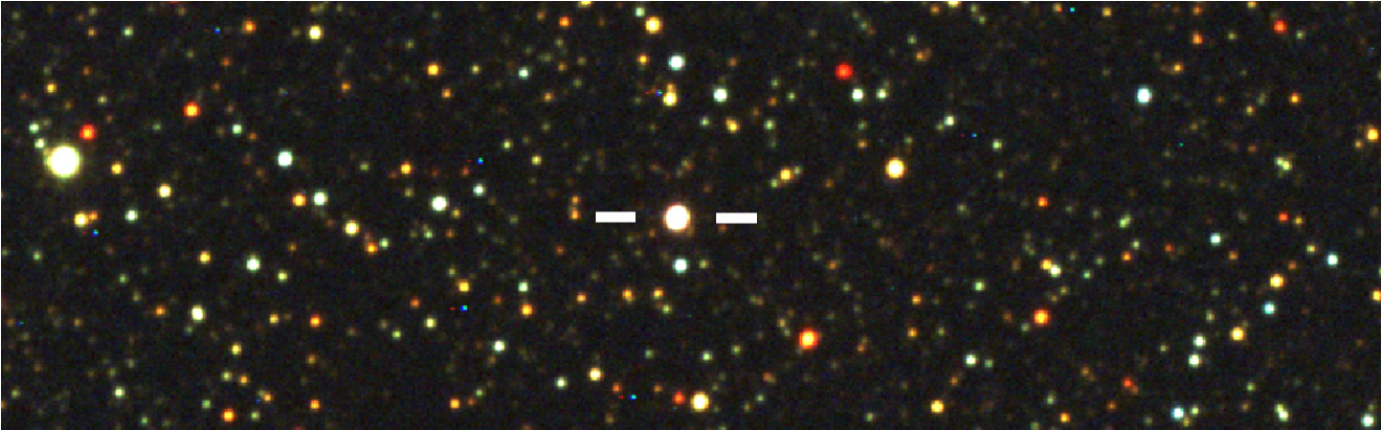

*Figure 14: Nova FQ Cir imaged remotely at iTelescope.Net T30 (0.50-m f/6.8 reflector in Siding Spring Observatory, Australia, MPC code Q62) on 2022-06-25.68: stacked 3x60 sec. with V, B and Ic filters for green, blue and red channels*

### 3.2. Observations of blazars

I regularly observe some blazars, for example, figure 15 shows the Rc light curve of BL Lac from September 2020 to April 2022 according to the results of my measurements from photographs obtained remotely at the telescopes of BGO and ARO. For some bright states of blazars, I wrote the ATels: #14467 «New peak of brightness of BL Lacertae» (*Romanov 2021*) 2021ATel14467....1R and #15399 «Follow-up optical photometry of flaring blazars S4 0954+65 and 1308+326 (AU CVn)» (*Romanov 2022*) 2022ATel15399....1R.

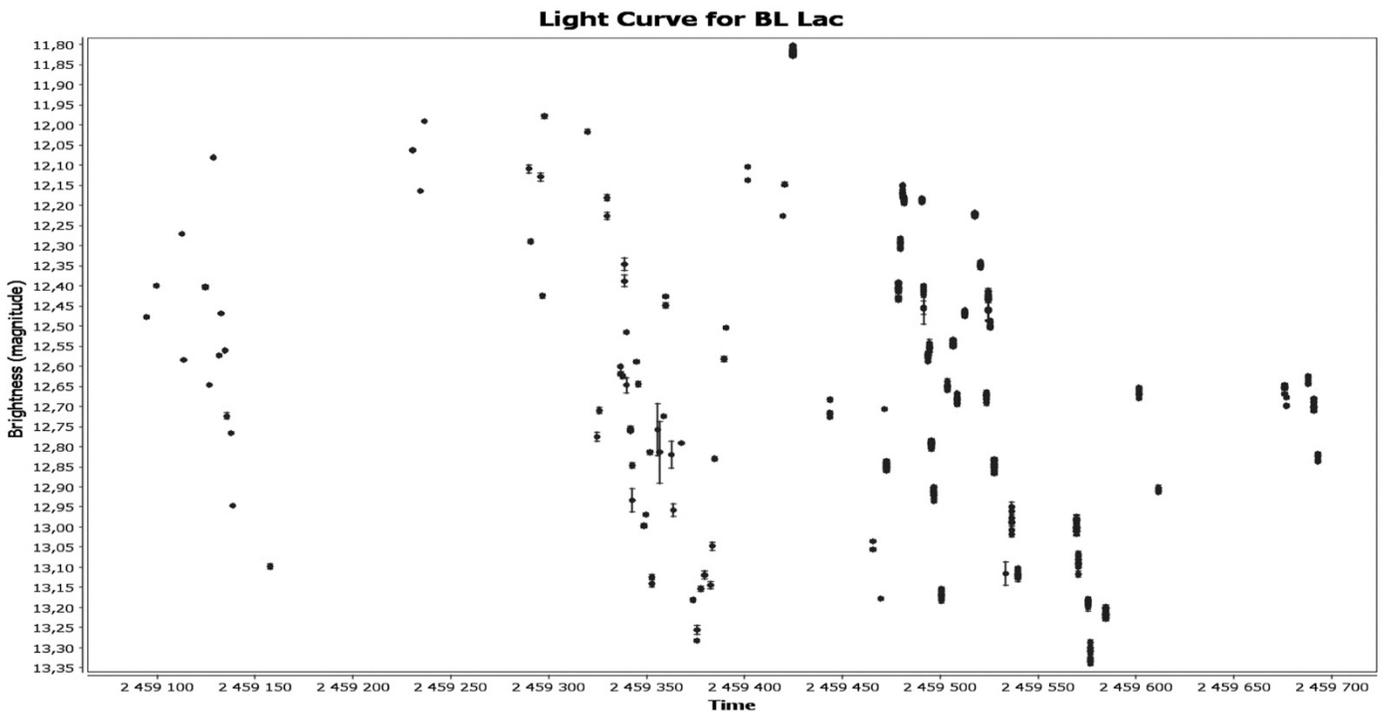

*Figure 15: JD light curve*

### 3.3.  Microlensing events

I took photometric measurements of several microlensing events, such as Gaia21efs: I observed it remotely using the ARO telescope with V, B, Rc and Ic filters. My V-band light curve shown in the figure 16 (from November 2021 to January 2022).

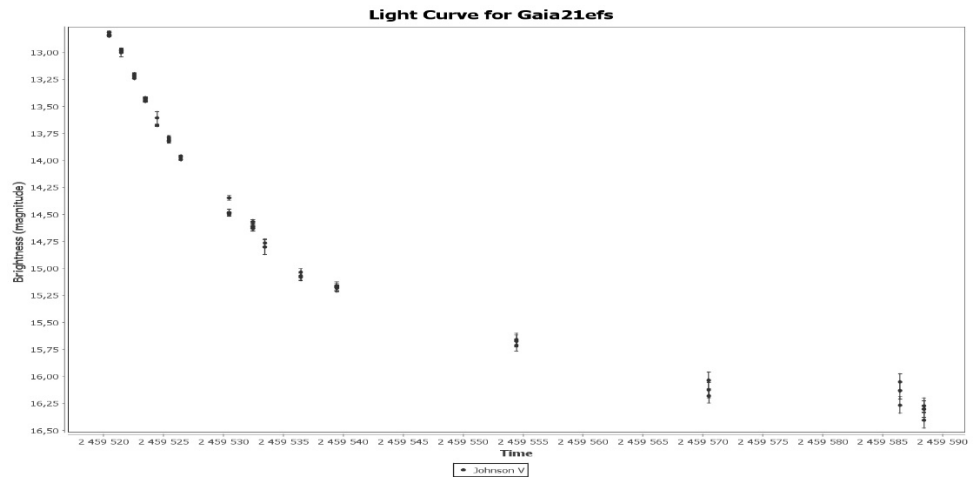

*Figure 16: JD light curve*

Also, in the ATel #14977 (*Romanov 2021*) 2021ATel14977....1R I presented my prediscovery photometric observations of the M33 microlensing event candidate AT 2021abdj from the images taken with the BGO telescope during the search for novae.

### 3.4.  Optical afterglows of gamma-ray bursts

I track information and make photometric measurements of them. I publish the results of my observations (brightness and positions) in the circulars of the Gamma-ray Coordinates Network (GCN). Examples: GRB 191221B (*Romanov 2019*) 2019GCN.26565....1R (figure 17) and GRB 210306A (*Romanov 2021*) 2021GCN.29599....1R; during the preparation of this paper, I observed the afterglow of record GRB 221009A (*Romanov 2022*) 2022GCN.32664....1R: in the figure 18.

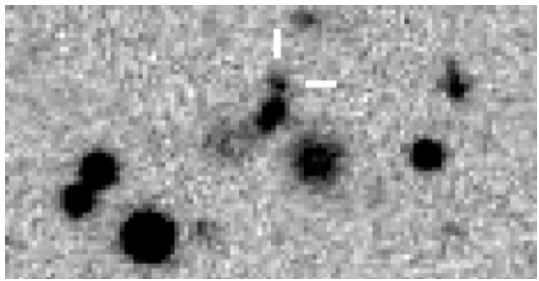

*Figure 17: Optical afterglow of GRB 191221B imaged on 2019-12-22 at iTelescope.Net T32 (0.43-m f/6.8 reflector + CCD) located in Siding Spring Observatory (Australia)*

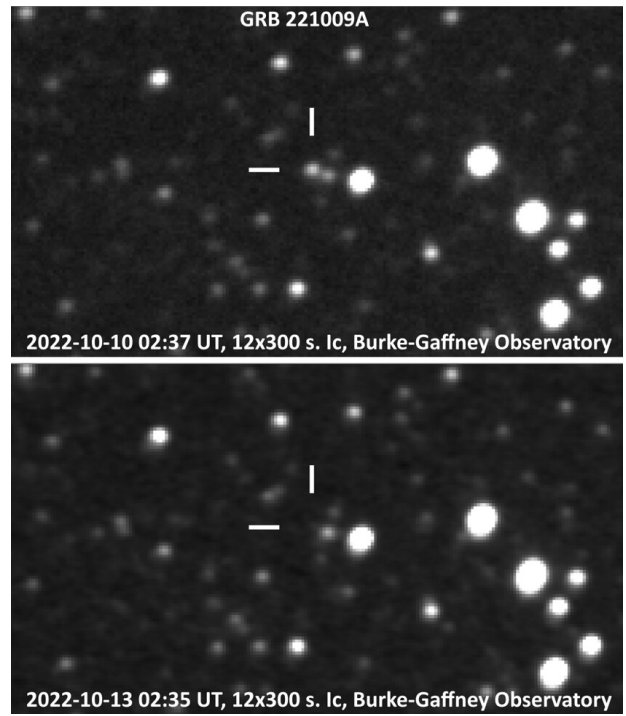

*Figure 18. Optical afterglow of GRB 221009A imaged at the BGO telescope with Ic filter*

### 3.5. Photometry of asteroid

In January 2022, remotely using the telescopes of ARO and BGO, I observed 2022 AB: fast rotating near-Earth asteroid. I made the phased light curve (using my photometric data from the ARO images, exposure times were 18 and 10 seconds for each image) with period = 181.87 sec. (*Kwiatkowski et al. 2023*, in preparation).

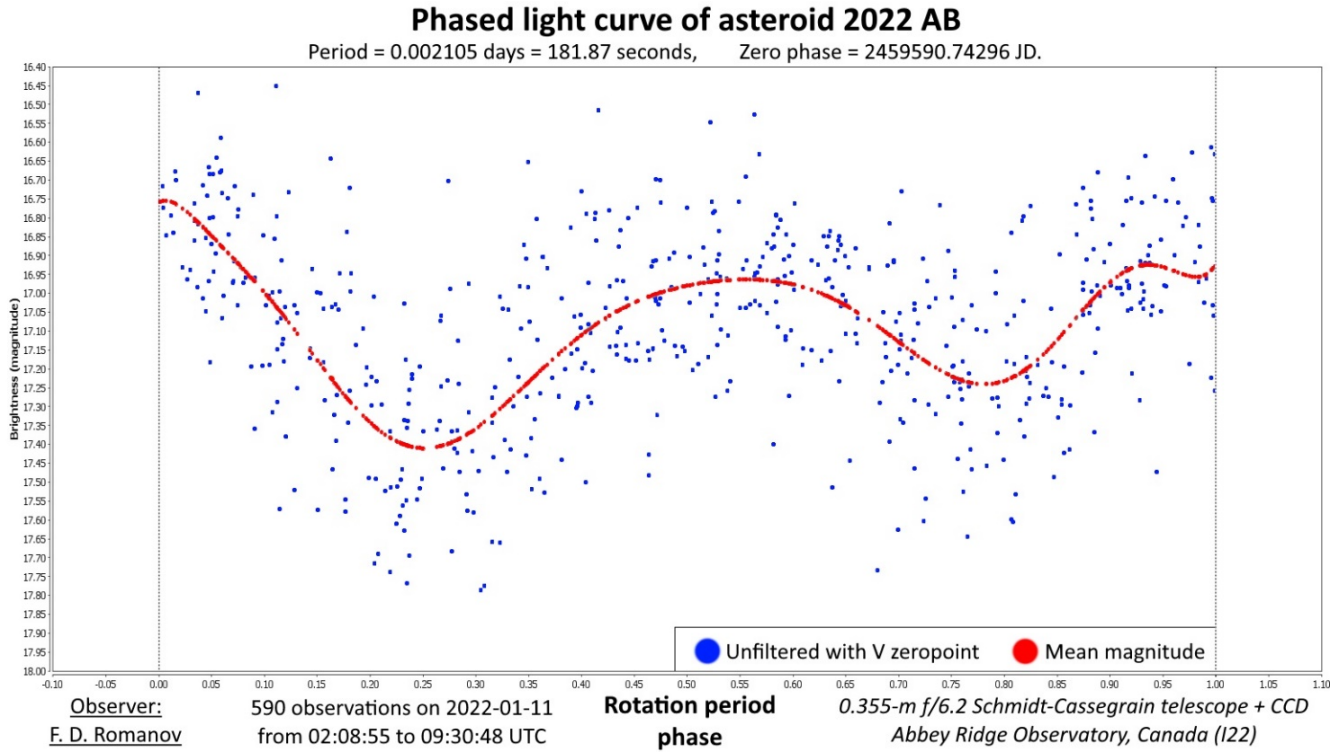

*Figure 19. For this light curve, I did a data analysis using VStar software (Benn 2012) 2012JAVSO..40..852B*

### 3.6. Astrometric observations of asteroids and comets

I confirm with my astrometric data the objects that are posted at the Confirmation Pages (NEO and PCCP) of the Minor Planet Center: asteroids (example: asteroid 2020 AV2 2020MPEC....A...99B (*Bolin et al. 2020*) = (594913) 'Aylo'chaxnim (the first in the class of interior to Venus asteroids: *Bolin et al. 2022* 2022MNRAS.517L..49B) - the photo taken using iTelescope.Net T21 on January 8, 2020, is presented in the figure 20) and comets (example: C/2021 A1 (Leonard) 2021MPEC....A...99L – *Leonard et al. 2021*), for some of which I also measure the diameter of the coma, the position angle and length of the tail.

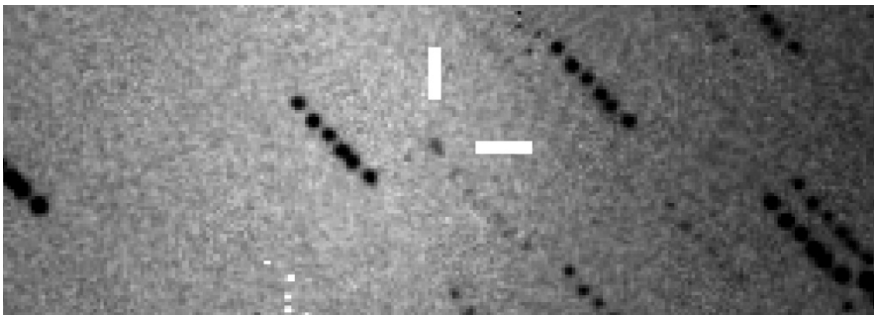

*Figure 20: stacked image (6x300 sec.) of 2020 AV2 from photos obtained on January 8, 2020, from 01:07 to 01:40 UT*

In November 2021, I took part in the timing campaign for asteroid 2019 XS organized by the International Asteroid Warning Network: I made astrometric measurements using the ARO telescope. The results of this have been published in the paper 2022PSJ.....3..156F (*Farnocchia et al. 2022*). I also contributed to the recovery of comet P/2007 R4 = P/2021 U2 (Garradd): details are published in the CBET 5061 (*Green 2021*).

I also made the astrometric measurements of interstellar comet 2I/Borisov, for example, my astrometric data has been published in the MPEC 2019-S62 : COMET C/2019 Q4 (Borisov) 2019MPEC....S...62B (*Bacci et al. 2019*): it was the last Minor Planet Electronic Circular (MPEC) before the interstellar designation of this comet.

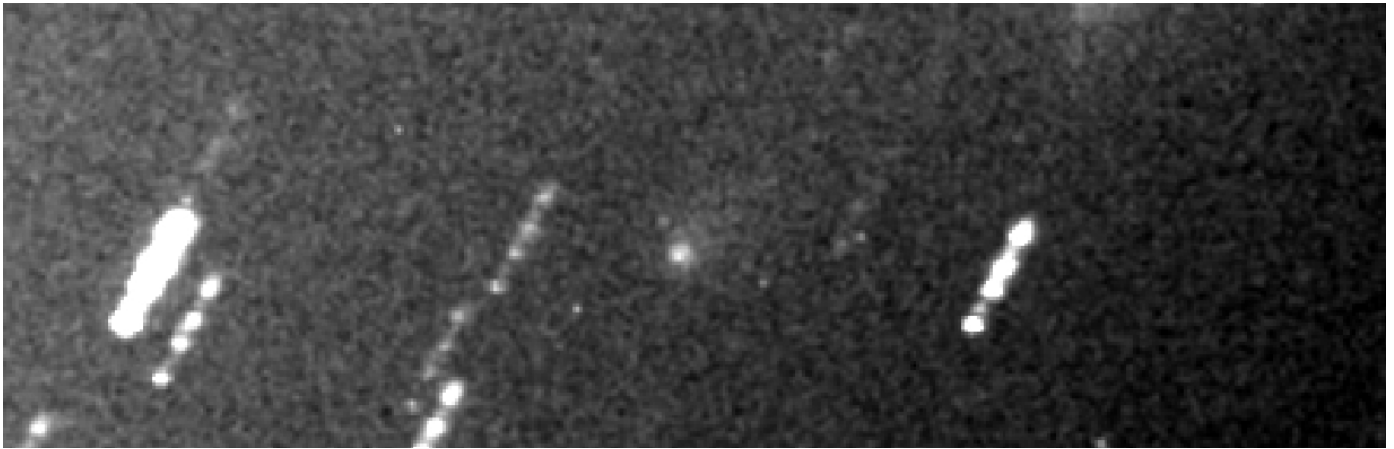

*Figure 21: comet 2I/Borisov imaged on December 8, 2019 (from 16:03 to 16:22 UT; during perihelion), remotely using iTelescope.Net T31 (0.50-m f/6.8 reflector + CCD + focal reducer at Siding Spring Observatory, Australia): stacked 6x120 sec. + 2x60 sec., Luminance filter*

During the preparation of this paper, I observed asteroid (65803) Didymos remotely at iTelescope.Net T69 (0.28-m f/2.2 reflector + CCD at Siding Spring Observatory, Australia) on 2022-09-29, near 16:44 UT, after the DART impact (*Cheng et al. 2018 2018P&SS..157..104C*). I detected three tails in the image (stacked 7x120 sec.) (figure 22):

- long, straight and bright: 170 arcseconds in length at PA = 290°;

- short, straight and faint: 90 arcseconds in length at PA = 130°;

- broad and faint: up to 70 arcseconds in length at PA = from 320° to 20°.

I estimate it was 1 magnitude brighter compared to September 17[th]: I observed it remotely at iTelescope.Net T32 near 2022-09-17.73 UT and near 2022-09-30.61 UT, I measured the magnitudes near 14.5 G and 13.5 G respectively (in comparison with Gaia DR2 magnitudes).

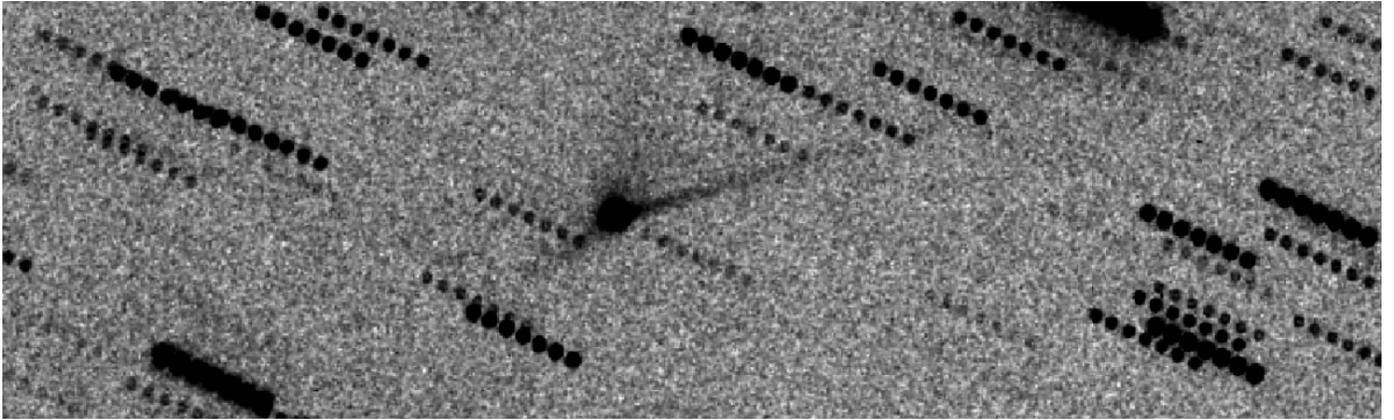

*Figure 22: image (changed from color to black and white) of asteroid (65803) Didymos that looks like «artificial comet»*

During the preparation of this paper, I found the asteroids [2022 WY16](#) and [2022 WY17](#) in the images obtained at the Liverpool Telescope according to my requests.

### 3.7. Astrometric observations of JWST

In addition to astrometry of comets and asteroids, I did astrometric observations of the James Webb Space Telescope (JWST) = 2021-130A = NORAD 50463 launched on December 25, 2021. Figure 23 shows stacked image (10x60 sec. with Luminance filter) from the photos obtained after the second mid-course correction burn, remotely at the BGO telescope. My astrometric data (made using Gaia DR2 catalogue: *Gaia Collaboration et al. 2018* [2018A&A...616A...1G](#); has been published in the list [https://projectpluto.com/pluto/mpecs/jwst_new.htm](https://projectpluto.com/pluto/mpecs/jwst_new.htm) on Bill Gray's website «Project Pluto»):

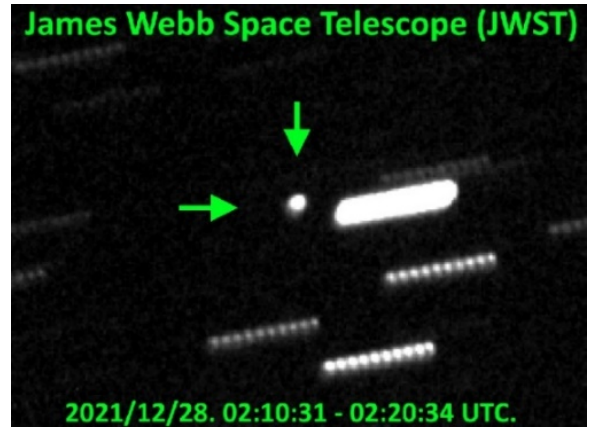

*Figure 23*

```
JWST01   C2021 12 28.09064 05 57 34.54 -00 36 31.3          15.3 G      851
JWST01   C2021 12 28.09141 05 57 34.25 -00 36 30.4          15.4 G      851
JWST01   C2021 12 28.09219 05 57 33.93 -00 36 29.1          15.5 G      851
JWST01   C2021 12 28.09296 05 57 33.64 -00 36 28.5          15.4 G      851
JWST01   C2021 12 28.09374 05 57 33.34 -00 36 27.5          15.5 G      851
JWST01   C2021 12 28.09451 05 57 33.03 -00 36 26.6          15.4 G      851
JWST01   C2021 12 28.09529 05 57 32.73 -00 36 25.9          15.3 G      851
JWST01   C2021 12 28.09606 05 57 32.42 -00 36 24.9          15.3 G      851
JWST01   C2021 12 28.09684 05 57 32.11 -00 36 24.0          15.3 G      851
JWST01   C2021 12 28.09762 05 57 31.80 -00 36 22.9          15.3 G      851
```

## 4. Observations of occultations

On 2016-09-01, when I still had the opportunity to live in my room in communal apartment in Moscow, I photographed the occultation of star TYC 6349-00855-1 by asteroid (159) Aemilia through my 0.2-m f/5 reflecting telescope (I presented this result at the International Occultation Timing Association's 39th Annual Meeting which was held online on July 17-18, 2021). Without access to my astronomical equipment, I can only observe at remote telescopes, for example, I photographed the occultation of SAO 164648 by Titan.

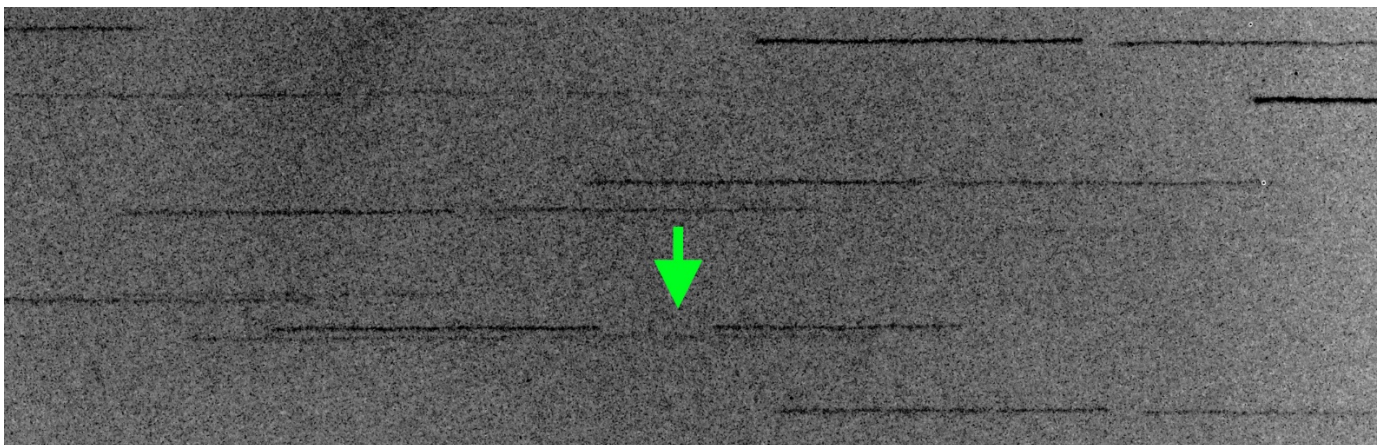

*Figure 24: In the image (two photos in a row with exposures of 30 sec.) taken (from 55.639361 N, 37.673139 E) near 19:55-56 UTC with Canon EOS 60D camera through telescope: reappearance of the star was detected (disappearance happened during 4 sec. interruption between photos) with duration > 7.5 sec.*

### 4.1. Occultation of the star by Saturn's moon Titan on 2022 July 9

Occultation of 8.7-mag star SAO 164648 = HIP 107569 = HD 207123 by Saturn's moon Titan was visible in North America. It was a rare opportunity to study Titan's atmosphere from Earth, and as an amateur astronomer, I was able to photograph this phenomenon. I used remote telescope T68 (of iTelescope.Net): 11" (280 mm) f/2.2 RASA astrograph with CMOS (ZWO ASI2600 Color) located in H06 observatory at Mayhill, New Mexico, USA. The weather forecast did not promise clear skies, but a few hours before the occultation the sky cleared up there. Immediately after the telescope has finished being used by another observer, I requested color images with exposures of 10 seconds for each.

The first obtained image (in the figure 25) shows the moons of Saturn: Rhea, Dione, Tethys and Titan (together with this star, which was at such a close angular distance that it could not be seen separately, but their contribution to the brightness was combined) before the occultation. On the second, the combined brightness of Titan and the star became less due to the fact that the star was occulted by the atmosphere of Titan. On the third, fourth, fifth (in the figure 25) and sixth images, the star was occulted: only the brightness of Titan was recorded, because the star was behind it. In the seventh image, the star was again shining through Titan's atmosphere, so their combined brightness began to increase. On the eighth image, occultation has already ended.

I extracted the images from the green channel and used the V magnitudes of the comparison stars from APASS DR9 (*Henden et al. 2016*) 2016yCat.2336....0H catalogue, (I also used the data from this catalog for photometry of 2022 AB) I did the photometric measurements and as a result I received TG magnitudes and made a light curve (figure 26). Based on these data, I determined that the total duration of occultation was within 259.64 and 339.3 seconds (started between 09:17:38.69 and 09:18:18.19 UTC, ended between 09:22:37.83 and 09:23:17.96 UTC); it is impossible to know more precisely, because the frame rate was limited. The time was synchronized with NTP server.

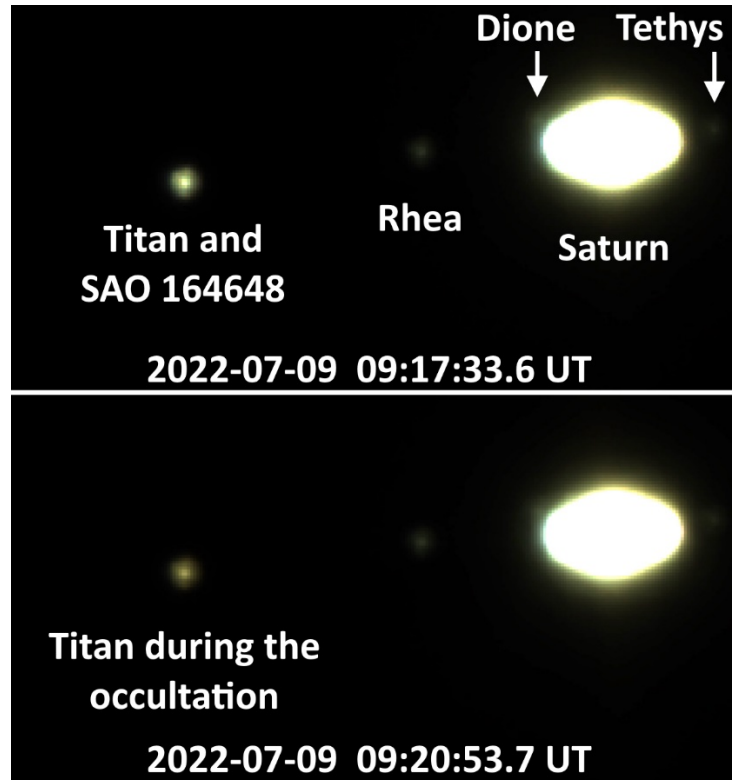

*Figure 25: 1st and 5th images*

*Table 1: Obtained data:*

| Image № | Time of the start of the exposure, UTC | Time of the end of the exposure, UTC | Mean JD | Magnitude, TG | Comment |
|---|---|---|---|---|---|
| 1 | 09:17:28.69 | 09:17:38.69 | 2459769.8871954861 | 8.04 | Occultation has not started yet |
| 2 | 09:18:18.19 | 09:18:28.19 | 2459769.8877684032 | 8.48 | The star was occulted by Titan's atmosphere |
| 3 | 09:19:08.81 | 09:19:18.81 | 2459769.8883542824 | 8.64 | The star was occulted by Titan |
| 4 | 09:19:59.11 | 09:20:09.11 | 2459769.8889364586 | 8.66 | The star was occulted by Titan |
| 5 | 09:20:48.72 | 09:20:58.72 | 2459769.8895106483 | 8.67 | The star was occulted by Titan |
| 6 | 09:21:39.09 | 09:21:49.09 | 2459769.8900936344 | 8.64 | The star was occulted by Titan |
| 7 | 09:22:27.83 | 09:22:37.83 | 2459769.8906577546 | 8.53 | The star was occulted by Titan's atmosphere |
| 8 | 09:23:17.96 | 09:23:27.96 | 2459769.8912379630 | 8.06 | Occultation has already ended |

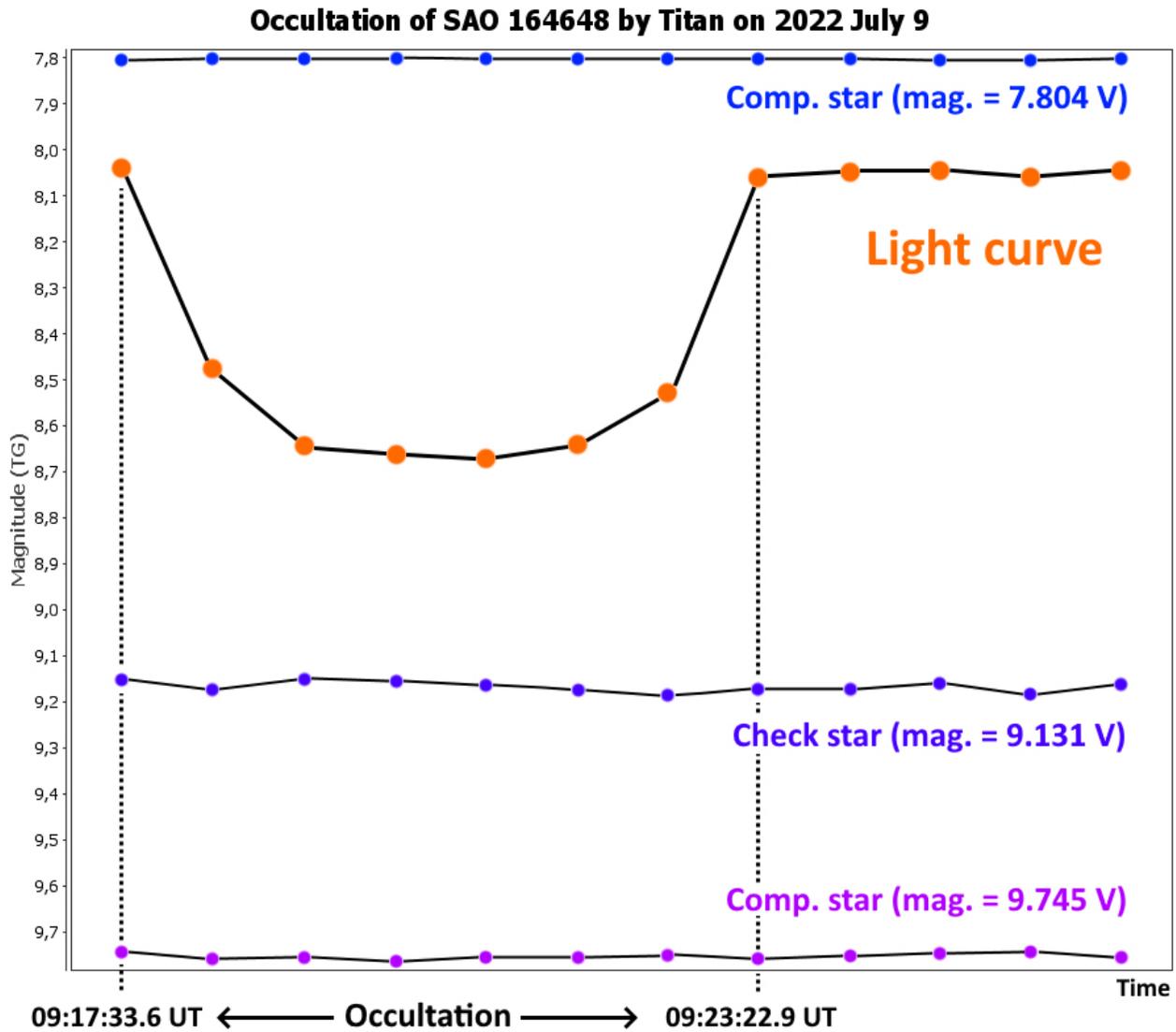

*Figure 26: Light curve*

## 4.2.  Occultation of Uranus by the Moon
### during total lunar eclipse on November 8, 2022

Occultation of a planet by the Moon during total lunar eclipse is a rare event that will next occur on June 2, 2235 (before this date, such occultations will occur during partial eclipses). During the preparation of this paper, I imaged this event on November 8, 2022, from Yuzhno-Morskoy, Nakhodka, Primorsky Krai, Russia (at 42.860200 N, 132.688762 E), in the apartment of my grandparents (access to this housing was given by my grandfather Viktor Nikandrovich Romanov: the father of my mother Larisa Viktorovna Romanova).

I used the telescope TAL-150K (150/1550 mm, provided by local amateur astronomer Dmitrii N. Grishin from Ussuriysk) and my camera Canon EOS 60D (with a broken display, so in order to see the image, I connected the camera with a cable to my laptop, and at the same time I was doing the live stream on YouTube: free for everyone, as a continuation of my popularization of astronomy, which I talked about in detail in my presentation during Communication Astronomy with the Public Conference 2021 which was held online on May 24th – 27th, 2021: *Romanov 2021* 2021cap..conf..114R): cropped video mode 640x480, ISO-6400, 1/30 sec. exposure time for each frame. I didn't have the technical ability to accurately detect the moments of time of the event.

In the figure 27: the first image (Moon and Uranus before the start of the occultation) was taken during the total phase of the eclipse (is the result of stacking of 332 frames from the video), a few minutes before the end of it. The second image (Moon and Uranus after the end of the occultation) was taken during the partial phase of the eclipse (is the result of stacking of 37 frames from the video), a few minutes before the end of it. Due to the fact that the start and end of the occultation occurred only for a few seconds, there were not so many frames for stacking, and there were few sharp ones among them. I stacked the frames using RegiStax 5.1 software 2012ascl.soft06001B (*Berrevoets et al. 2012*).

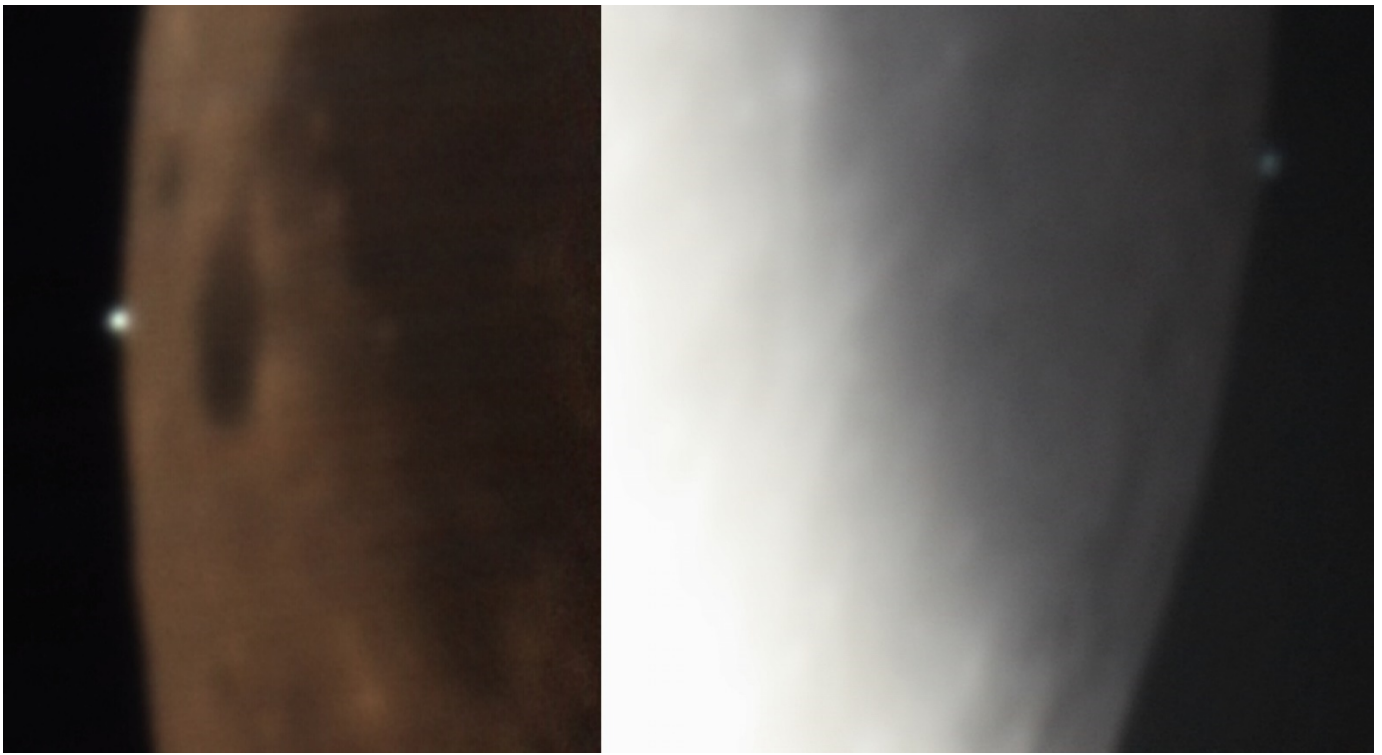

*Figure 27*

I was waiting for this phenomenon for about 10 years, and the weather was very clear, but an hour after the end of the penumbral phases of the eclipse, the sky was covered with clouds. The last time before this I observed (from 42.870555 N, 132.67444 E) and photographed a total lunar eclipse on the night of December 10-11, 2011 (that is, with a difference of one

Tritos eclipse cycle). Year before, on November 19, 2021, from Yuzhno-Morskoy, I observed and photographed the partial lunar eclipse (by the way, one Tritos cycle before that, I observed and photographed the total lunar eclipse on December 21, 2010).

### 5. Visual observations

I contribute to the science also through my visual observations:

I observed Nova Cas 2021 with the unaided eye on 2021-05-10 at 18:00 UT from my small homeland: from Yuzhno-Morskoy, Nakhodka, Primorsky Krai, Russia (at 42.858127 N, 132.687517 E). It was easily seen next to the star 4 Cas against the background of the Milky Way. I estimated its brightness as 5.1 mag during this observation (_Romanov 2021_ [vsnet-alert 25848]).

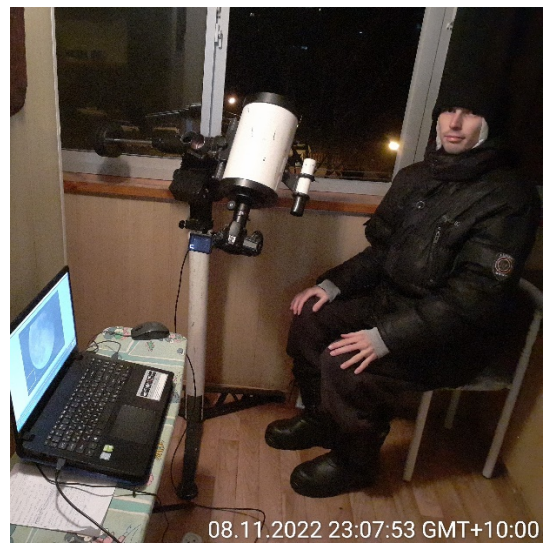

*Figure 28: After the end of the partial phases of the eclipse on 2022-11-08 and after the end of the observations of occultation of Uranus by the Moon. Polar alignment was not done due to space limits*

Also, from Yuzhno-Morskoy I observed RS Oph with the unaided eye on 2021-08-09 at 15:20 UT. I have been waiting for several hours for gaps in the clouds, and the sky cleared for a few minutes: it was easily seen in the sky and I estimated its brightness as 4.6 mag (_Romanov 2021_ [vsnet-alert 26141]).

Another example: during a year I submitted my visual observations (that were made every clear day) of sunspots to the AAVSO database. Even before that, since 2010, when I had not yet sent the results of my observations to the AAVSO, I observed sunspots, counted and sketched them and photographed them through my telescopes, including even during solar eclipses on May 21, 2012 (from Yuzhno-Morskoy) and March 20, 2015 (from Korolyov, Moscow Oblast), and during the transit of Venus on June 6, 2012 (from Ussuriysk Astrophysical Observatory: UAO) and during the Mercury transit on May 9, 2016 (from Moscow).

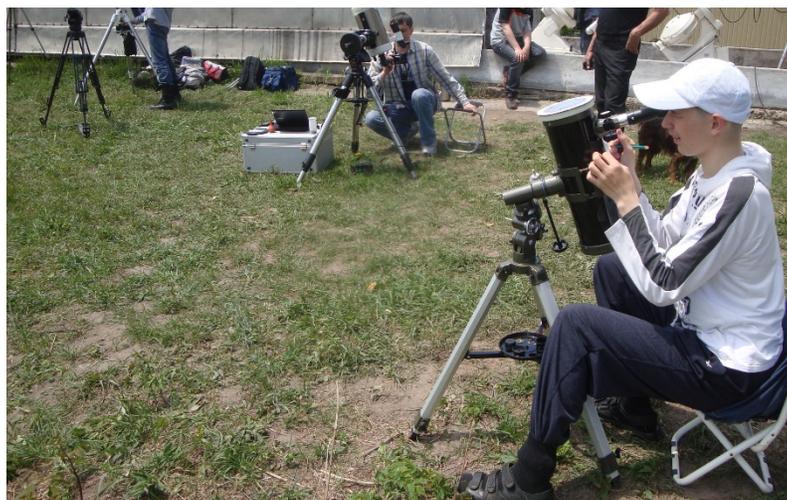

*Figure 29: On the territory of the UAO, during the observation of the transit of Venus, 2012-06-06, 02:44 UTC*

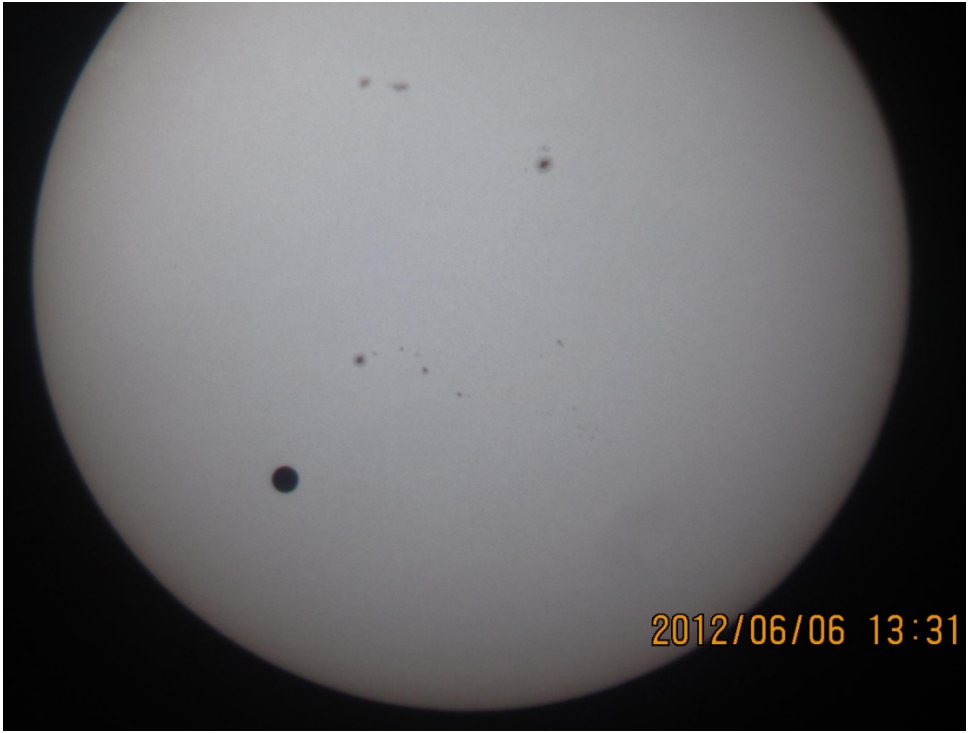

*Figure 30. Transit of Venus on 2012-06-06 (last until 2117) and sunspots imaged by me on 02:31 UTC with my camera Canon IXUS 310 HS and through my 114/1000 mm reflector with a solar filter - from the territory of UAO (43.698579 N, 132.165790 E) when the sky was completely cleared. The planet was clearly visible in front of the solar disk even without optics, with the unaided eye through the solar filter. Inverted image*

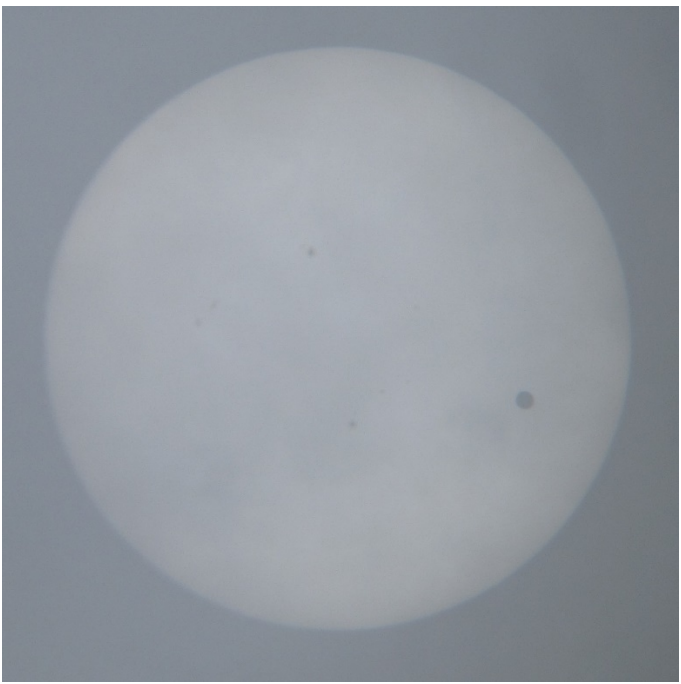

*Figure 31. I did not see the start of transit of Venus because it was cloudy and foggy, but at 11:08 local time (00:08 UTC) I took this photo (through my 114-mm telescope, **without solar filter**) of the transit the first among the people (amateur and professional astronomers, journalists and interested people) who were present there, during the hole in the clouds*

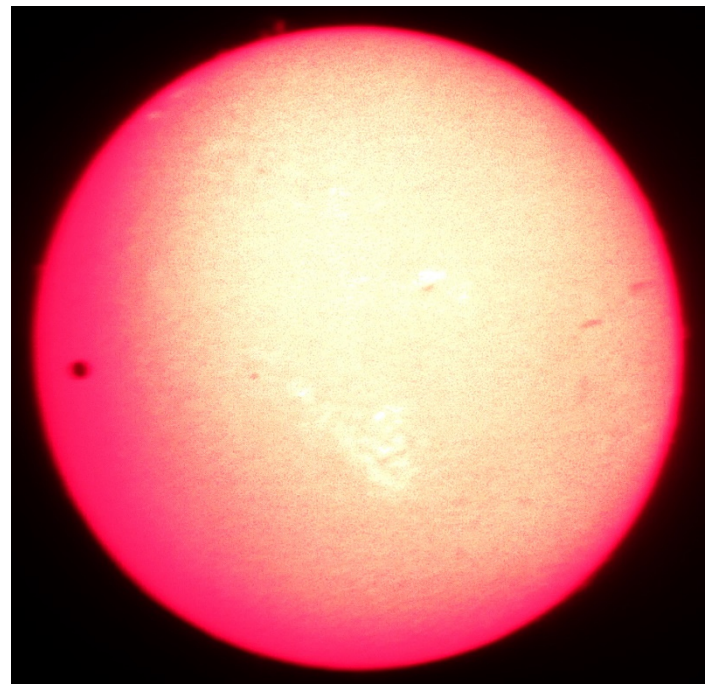

*Figure 32. Then the sky cleared and I saw the middle and end of the transit. I did not see the atmosphere of Venus, but I looked at the Sun through the Coronado P.S.T. H-alpha solar telescope (which belonged to one of amateur astronomers) for the first time in my life and took this photo on 04:01 UTC*

Since 2013, I have been regularly observing meteor showers (Perseids, Lyrids, Southern Delta Aquariids, Alpha Capricornids, Geminids, Draconids, Eta Aquarids, Taurids, Orionids, Ursids, Quadrantids) visually (I estimate the brightness and direction of meteors, detect time) and send data to the International Meteor Organization, reports are available in my profile.

In my paper 2021JIMO...49..158R (*Romanov 2021*) I reported my visual and photographic (using DSLR camera) observations of the Geminids for different years: from 2016 to 2020 (I presented this my work at the 40th International Meteor Conference which was held online on September 25th/26th, 2021).

## 6. Conclusion

In this paper, I showed by my example that if an amateur astronomer is really interested and has access to quality astronomical equipment, as well as appropriate skills, then amateur astronomer can study astronomical objects, make various astronomical observations and contribute to science. I hope to make more of my contribution to the science in future years, including already as a professional astronomer after receiving an astronomical education.

## Acknowledgements

I am grateful to iTelescope.Net for giving me some complimentary points for observing time with their remote telescopes.

I am also grateful to David J. Lane for allowing me to use the remote robotic telescope of his private Abbey Ridge Observatory for obtaining scientific data.

I am grateful to Liverpool John Moores University for giving me observation time at the Liverpool Telescope after my proposal through the PATAG program. The Liverpool Telescope is operated on the island of La Palma by Liverpool John Moores University in the Spanish Observatorio del Roque de los Muchachos of the Instituto de Astrofisica de Canarias with financial support from the UK Science and Technology Facilities Council.

I am grateful to the AAVSO for giving me a complimentary membership for 2022.

I am grateful to Prof. Nancy Morrison (Editor-in-Chief of The Journal of the American Association of Variable Star Observers) for giving the endorsement letter for my KAS grant application.

I am grateful to the Korean Astronomical Society for giving me the KAS Grant for my complimentary virtual participation (as virtual attendee with e-Poster) in the IAUGA 2022.

# References

Bacci, P., et al. 2019, *Minor Planet Electronic Circulars* No. **2019-S62**

Berrevoets, C., et al, 2012, *Astrophysics Source Code Library*, record ascl:1206.001

Benn, D. 2012, *The Journal of the American Association of Variable Star Observers*, vol. **40**, no. 2, p. 852

Bolin, B. T., et al. 2020, *Minor Planet Electronic Circulars* No. **2020-A99**

Bolin, B. T., et al. 2022, *Monthly Notices of the Royal Astronomical Society: Letters*, Volume **517**, Issue 1, November 2022, Pages L49–L54

Chambers, K. C., et al. 2016, *arXiv e-Print*: **1612.05560**

Cheng, A. F., et al. 2018, *Planetary and Space Science*, Volume **157**, p. 104-115

Farnocchia, D., et al. 2022, *The Planetary Science Journal*, Volume **3**, Issue 7, id.156

Fremling, C., Romanov, F. and Zwicky Transient Facility collaboration, Transient Name Server Classification Report, No. **2022-553**

Gaia Collaboration, et al. 2018, *Astronomy & Astrophysics*, **616A**, 1.

Green, D. W. E. 2021, *Central Bureau Electronic Telegrams* No. **4945**

Green, D. W. E. 2021, *Central Bureau Electronic Telegrams* No. **4976**

Green, D. W. E. 2021, *Central Bureau Electronic Telegrams* No. **4977**

Green, D. W. E. 2021, *Central Bureau Electronic Telegrams* No. **5061**

Henden, A. A., et al. 2016, *VizieR On-line Data Catalog*: II/336

Isogai, K., et al. 2021, *The Astronomer's Telegram*, No. **15074**

Kato, T. and Romanov, F. D. 2022, *VSOLJ Variable Star Bulletin* No. 98, p. 1

Kwiatkowski, T., et al. 2023, the paper is in preparation

Lane D. J., 2018, *Proceeding of the Robotic Telescopes, Student Research, and Education Conference*, Vol. **1**, No. 1, pp. 119-126

Le Du, P. 2019, *VizieR Online Data Catalog: J/other/LAstr/125.54*

Leonard, G. J., et al. 2021, *Minor Planet Electronic Circulars*, No. **2021-A99**

Ochner, P., et al. 2022, *Transient Name Server Classification Report*, No. **2022-2436**

Parker, Q. A., et al. 2005, *Monthly Notices of the Royal Astronomical Society*, Volume **362**, Issue 2, pp. 689-710

Perley, D., et al. 2022, *Transient Name Server Classification Report*, No. **2022-2345**

Rivera Sandoval, L. E., et al. 2022, *The Astrophysical Journal*, Volume **926**, Issue 1, id.10

Romanov, F. D. 2018, *Open European Journal on Variable stars*, Vol. **190**, p. 1

Romanov, F. D. 2019, *Journal of Double Star Observations*, vol. **15**, no. 3, p. 434-435

Romanov, F. 2019, *Transient Name Server Discovery Report*, No. **2019-2353**

Romanov, F. 2019, *Transient Name Server Discovery Report*, No. **2019-2388**

Romanov, F. D. 2019, *GRB Coordinates Network, Circular Service*, No. **26565**

Romanov, F. 2020, *Transient Name Server Discovery Report*, No. **2020-2387**

Romanov, F. D. 2021, *GRB Coordinates Network, Circular Service*, No. **29599**

Romanov, F. 2021, *The Astronomer's Telegram*, No. **14467**

Romanov, F. 2021, *VSNET Alert* No. **25848**

Romanov, F. 2021, *VSNET Alert* No. **26141**

Romanov, F. 2021, *The Astronomer's Telegram*, No. **14977**

Romanov, F. 2021, *Book of Proceedings Communication Astronomy with the Public Conference 2021*, p. 114

Romanov, F. 2021, *WGN, Journal of the International Meteor Organization*, vol. **49**, no. 6, p. 158-162

Romanov, Filipp Dmitrievich 2021, *The Journal of the American Association of Variable Star Observers*, vol. **49**, no. 2, p. 130

Romanov, F. and CRTS 2022, *Transient Name Server Discovery Report*, No. **2022-339**

Romanov, F. 2022, *The Astronomer's Telegram*, No. **15339**

Romanov, F. and CRTS 2022, *Transient Name Server Discovery Report*, No. **2022-1171**

Romanov, F. and Pastorello, A. 2022, *Transient Name Server Classification Report*, No. **2022-1273**

Romanov, F. 2022, *The Astronomer's Telegram*, No. **15399**

Romanov, F. 2022, *The Astronomer's Telegram*, No. **15511**

Romanov, F. 2022, *Transient Name Server Discovery Report*, No. **2022-1975**

Romanov, F. 2022, *Transient Name Server Discovery Report*, No. **2022-2248**

Romanov, F. 2022, *Transient Name Server Discovery Report*, No. **2022-2276**

Romanov, F. 2022, *Transient Name Server Classification Report*, No. **2022-2408**

Romanov, F. 2022, *The Astronomer's Telegram*, No. **15569**

Romanov, F. D. 2022, *GRB Coordinates Network, Circular Service*, No. **32664**

Steele, I. A., et al. 2004, *Ground-based Telescopes. Edited by Oschmann, Jacobus M., Jr., Proceedings of the SPIE*, Vol. **5489**, pp. 679-692

Watson, C. L., Henden, A. A., and Price, A. 2006, in *The Society for Astronomical Sciences 25th Annual Symposium on Telescope Science*, Society for Astronomical Sciences, Rancho Cucamonga, CA, 47

Wright, E. L., et al. 2010, *The Astronomical Journal,* Volume **140**, Issue 6, pp. 1868-1881